\documentclass[amsmath,amssymb,aps,twocolumn,showkeys
]{revtex4-2}
\usepackage{tabularray}
\usepackage{booktabs}
\usepackage{tabularx}
\usepackage{booktabs}
\usepackage{graphicx}
\usepackage{dcolumn}
\usepackage{bm}
\usepackage{physics}
\usepackage{amsmath}
\usepackage{amsfonts}
\usepackage{xcolor}
\usepackage{hyperref}
\usepackage{url}
\usepackage{subfig}
\usepackage[font=small,labelfont=bf,justification=justified,format=plain]{caption}
\usepackage{multirow}

\begin{document}


\title{Harnessing Quantum Support Vector Machines for Cross-Domain Classification of Quantum States}

\author{Diksha Sharma$^{1}$}
\email{sharma.49@iitj.ac.in}
\author{Vivek Balasaheb Sabale$^{1}$}
\email{sabale.1@iitj.ac.in}
\author{Parvinder Singh$^{2}$}
\email{parvinder.singh@cup.edu.in}
\author{Atul Kumar$^{1}$}
\email{atulk@iitj.ac.in}
\affiliation{$^{1}$ Indian Institute of Technology Jodhpur, 342030, India}
\affiliation{$^{2}$Central University of Punjab Bathinda, Punjab-151001, India}

\date{\today}

\begin{abstract}
 
In the present study, we use cross-domain classification using quantum machine learning for quantum advantages to readdress the entanglement versus separability paradigm. The inherent structure of quantum states and its relation to a particular class of quantum states are used to intuitively classify testing states from domains different from training states, called \textit{cross-domain classification}. Using our quantum machine learning algorithm, we demonstrate efficient classifications of two-qubit mixed states into entangled and separable classes. For analyzing the quantumness of correlations, our model adequately classifies Bell diagonal states as zero and non-zero discord states.  In addition, we also extend our analysis to evaluate the robustness of our model using random local unitary transformations. Our results demonstrate the potential of the quantum support vector machine for classifying quantum states across the multi-dimensional Hilbert space in comparison to classical support vector machines and neural networks.
\end{abstract}
 

\maketitle

\section{\label{sec:level1}Introduction}
The entangled quantum states are known for the non-local correlations, defying the notion of classical mechanics confined by locality, and serve as the foundational block in quantum protocols like teleportation \cite{zeilinger2000quantum,zeilinger2018quantum,popescu1994bell,mehic2020quantum}, dense coding \cite{bennett1992communication,hwang2011quantum}, and secure key distribution \cite{bennett1988privacy,ekert1991quantum}. Apart from the quantum protocols, entangled states are also utilized in quantum networking \cite{van2014quantum,munro2010quantum}, quantum machine learning (QML) \cite{cai2015entanglement,biamonte2017quantum,sharma2023role}, quantum finance \cite{lee2020quantum, schaden2002quantum,orus2019quantum}, and variational quantum eigensolvers (VQE) \cite{tilly2022variational,kandala2017hardware,cerezo2022variational}.\par
In general, quantum states are categorized or classified into entangled and separable classes. With an increase in the size of the system, the potential classes for quantum state classification also increase. For example, bipartite or two-qubit quantum states are typically classified into two classes, while this classification expands to five classes for the three-qubit scenario \cite{PhysRevA.61.042314}. The intricacies further increase as we deal with bipartite and multi-qubit mixed states. Therefore, despite being a crucial and advantageous property of quantum computing, entanglement detection, identification, and classification tasks are extremely hard with the increasing number of sub-systems or qubits in the context of quantum information. \par
To address the entanglement versus separability paradigm, entanglement classification is approached by machine learning algorithms as a pattern recognition task. In one such attempt, Harney \textit{et al.} demonstrated the use of artificial neural networks and reinforcement learning to classify multi-qubit pure quantum states \cite{harney2020entanglement}. Asif \textit{et al.} also made use of an artificial neural network featuring multiple Bell-type inequalities for relative entropy of coherence to detect quantum states as separable or entangled states \cite{asif2023entanglement}. In addition to artificial neural networks, \cite{vintskevich2023classification} used support vector machine learning algorithms and entanglement witness operators to facilitate the classification of four-qubit arbitrary pure states. In \cite{lu2018separability}, authors show the efficacy of machine learning tools with bagging-based models to design an entanglement classifier using a convex hull of separable states. Moreover, the model also employs prior knowledge about the states, which further assists in classifying the states as separable or entangled. For an effective classification, machine learning algorithms require a large dataset for training and subsequently use a subset of the original dataset for testing. As the number of qubits increases in a quantum state, the dimensions of the density matrix- representing the features- increase exponentially in accordance with the associated Hilbert space. For example, a density operator representing an $n$-qubit quantum state is associated with $2^{n}$ basis states and $2^n \times 2^n$ features in the realms of classical machine learning.  To accurately classify the test states, classical machine learning algorithms require a large dataset spanning a complete vector or Hilbert space, consequently increasing the computational training time with increasing size of the system or increasing number of qubits.  \par
\begin{figure*}[htpb]
\captionsetup[subfigure]{labelformat=empty}
\subfloat[A. Visualization of the Werner type states]{\includegraphics[width=0.45\textwidth]{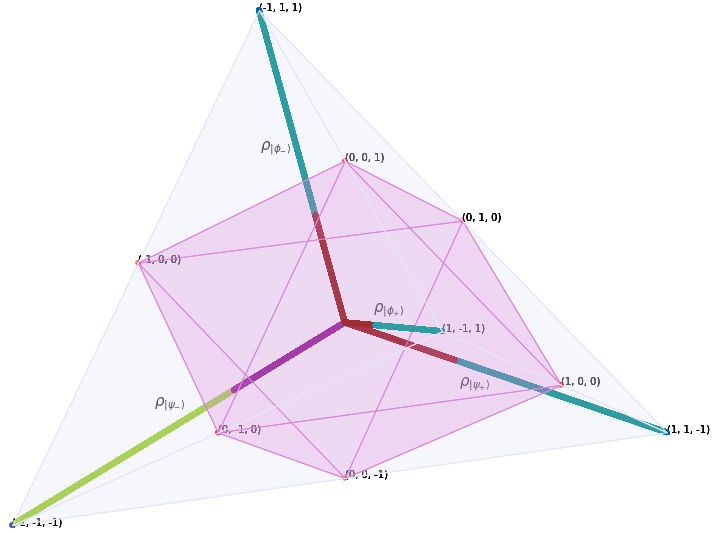}}
\subfloat[B. Randomly generated Bell diagonal states]{\includegraphics[width=0.45\textwidth]{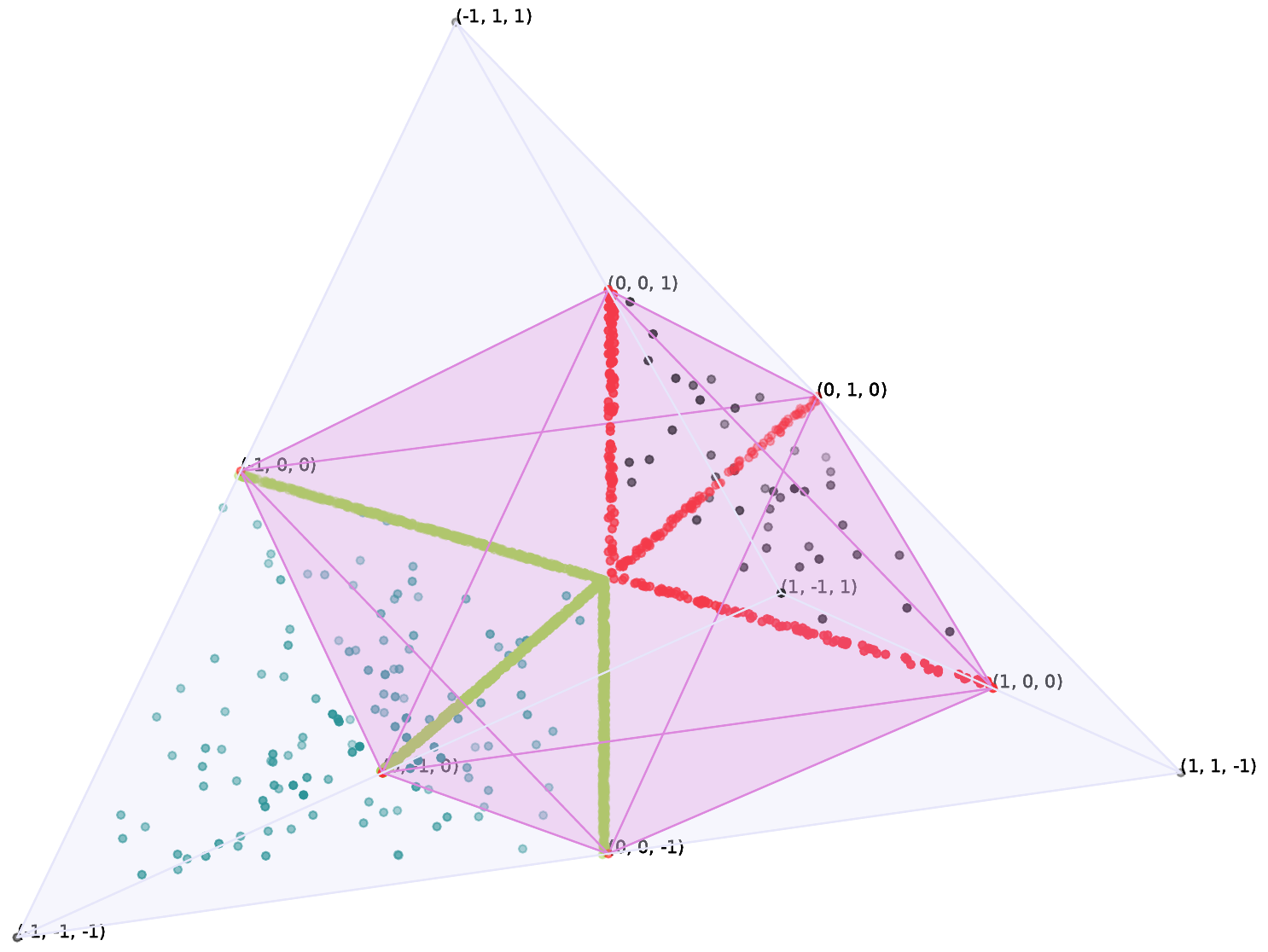}}
\centering
\captionof{figure}{A geometrical representation of training and testing domains for two-qubit mixed states- the four corners represent Bell states such that (-1,-1,-1), (-1,1,1), (1,-1,1), and (1,1,-1) represent $\ket{\psi_{-}}$, $\ket{\phi_{-}}$, $\ket{\phi_{+}}$, and $\ket{\psi_{+}}$, respectively. A) The line joining from the center to the vertex represents one of the Werner states, and the part of the line confined in the octahedron depicts the separable class. The part of the line confined outside the octahedron depicts the entangled class. B) For classifying the quantumness of correlations, the inside edges (red and green lines) show zero-discord states; all other states represent non-zero discord states.}
\label{fig:visualization}
\end{figure*}
In this article, we readdress the entanglement versus separability paradigm using quantum machine learning algorithms for \textit{cross-domain classification} where the realms of testing and training datasets are always different \cite{caro2023out,liu2021unified}. Our approach uses the inherent mathematical structure of quantum states as the basis for the optimal classification. Considering cross-domain classification, our analysis further facilitates reducing the size of a training set. Our algorithm effectively exploits the essence of entanglement and separability in training and testing states, which conventional methods do not recognize otherwise. 
In order to facilitate the discussion on cross-domain classification, let us consider the class of two-qubit mixed Werner states \cite{werner1989quantum}, namely
\begin{equation}\label{eq:Werner state}
    \rho_{\ket{\psi}}= p \ket{\psi}\bra{\psi} + \frac{(1-p)}{4}I,
\end{equation}
where, $\ket{\psi}$ represents one of the Bell states given as $\ket{\psi_{\pm}}=\frac{1}{\sqrt{2}}\ket{00}\pm \ket{11}$ or $\ket{\phi_{\pm}}=\frac{1}{\sqrt{2}}\ket{01}\pm \ket{10}$ and $p$ represents the probability. Fig. \ref{fig:visualization}A represents different types of Werner states where the straight lines, starting from the center of the cube and ending at the vertex of the tetrahedron, represent a type of Werner states.
Moreover, the states lying inside the octahedron belong to the separable class of Werner states; otherwise, the states belong to the entangled class \cite{YAO2012358,PhysRevLett.105.150501}. For the in-domain classification, both training and testing states belong to the same type of Werner states. In contrast to in-domain classification, for cross-domain classification, we train our model on one type of Werner state and test the model on other types of Werner states. We further analyze the efficacy of our model by testing it with Horodecki and maximally entangled mixed states (MEMS) states \cite{horodecki2009quantum,wei2003maximal,singh2018correlations,ishizaka2000maximally}, which are different from Werner states. On similar lines, for classifying states as zero discord or non-zero discord states in the cross-domain classification regime, one can use random Bell diagonal states \cite{PhysRevLett.105.150501} of the form 
\begin{equation}\label{eq:bd_state}
    \rho_{BD} = \frac{1}{4} (I \otimes I + \sum_{i=1}^{3} t_{ii} \sigma_{i} \otimes \sigma_{i} ),
\end{equation}
where $t_{ii}$ ranges from -1 to 1. For classification, states with zero discord constitute one of the classes, and states with non-zero discord constitute the other class. Fig. \ref{fig:visualization}B depicts the distinct difference in the cross-domain data with the range of $t_{ii}$ from -1 to 0 for training purposes and from 0 to 1 for testing. For classifying the quantumness of correlations, the inside edges (red and green lines) show zero-discord states; all other states represent non-zero discord states.  \par
Considering the complex natures of mixed states, the classification of such states as entangled and separable is a challenging yet an interesting problem. In this study, we use a quantum machine learning algorithm- in particular, a quantum support vector machine (QSVM)- having a fully entangled \textit{ZZ+UC}-gates-based circuit \cite{sharma2023role}. The purpose is to showcase the advantages of quantum algorithms in effectively recognizing patterns compared to their classical counterparts. For this, we first utilize our model for in-domain classification of two-qubit mixed states into separable and entangled classes and then present the analysis for training the QSVM on one type of Werner states $(\rho_{\ket{\psi_{-}}})$ and subsequently test the trained model with another type of Werner states $(\rho_{\ket{\psi_{+}}}), (\rho_{\ket{\phi_{-}}})$ and $ (\rho_{\ket{\phi_{+}}})$. Clearly, the training set belongs to one part of the Hilbert space, and the testing set belongs to a completely different part of the Hilbert space, as demonstrated in Fig. \ref{fig:visualization}. For a detailed analysis, we further use our trained model to classify Horodecki and MEMS. Our results demonstrate that the model can accurately predict testing states with a significant prediction probability. We also evaluate our algorithm based on different evaluation metrics, i.e., accuracy, precision, recall, and F1-score, and show that the model performs exceedingly well on all evaluation criteria. Interestingly, our model classifies Horodecki states as entangled or separable with $100\%$ accuracy. In addition, we thoroughly evaluate our model by performing random unitary operations on Werner states to understand its robustness and observe similar performance. For analyzing nonclassical correlations, we further use our  QSVM model to identify zero and non-zero discord states with a high success probability. Our model detects quantum correlations in different types of Werner states and classifies them as zero or non-zero discord states with significant accuracy. For a comparative analysis, we evaluate the CSVM and neural networks with the same strategies and observe that for cross-domain classification, the classical models predict each state of the testing set as separable states. 
\section{Preliminaries}
In this section, we present a brief overview of necessary concepts such as entanglement and quantum correlation measures for identifying the entangled and separable classes. The section also gives an introduction to the quantum machine learning algorithm, QSVM, used in this study. 
\subsection{Entanglement and quantum correlation measures}
One of the most fundamental and promising quantifiers of the degree of entanglement in two-qubit systems is concurrence \cite{PhysRevLett.80.2245}. The concurrence of a two-qubit quantum state $\rho$ is defined as
\begin{equation}
    C(\rho)=max\{ 0, \lambda_{1}-\lambda_{2}-\lambda_{3}-\lambda_{4}\},
\end{equation}
where $\lambda_{i}'s$ are square roots of eigenvalues of the matrix $\rho \Tilde{\rho}$ such that $\Tilde{\rho}=(\sigma_{y}\otimes\sigma_{y})\rho^{*}(\sigma_{y}\otimes\sigma_{y})$, and $\rho^{*}$ is complex conjugate of the state $\rho$. Similarly, geometrical discord- a signature of quantumness of correlations- can be used to detect nonclassical correlations in two-qubit states \cite{YAO2012358}. For a general two-qubit state, represented as 
\begin{equation}\label{eq:gdm}
    \rho = \frac{1}{4} (I \otimes I+ \sum_{i=1}^{3} a_{i}\sigma_{i}\otimes I+  \sum_{j=1}^{3} b_{j} I \otimes \sigma_{j} + \sum_{i,j=1}^{3} t_{ij} \sigma_{i} \otimes \sigma_{j} ),
\end{equation}
where $a_i = tr (\rho (\sigma_{i}\otimes I))$ and $b_j=tr (\rho (I \otimes\sigma_{j}))$ are individual polarization vectors and $t_{ij} = tr (\rho (\sigma_{i}\otimes\sigma_{j}))$ are elements of a correlation matrix, the geometrical discord is defined as 
\begin{equation}
    D_{G}(\rho)= \frac{1}{4}(||\Vec{a}||^{2} + ||T|| - \lambda_{max}),
\end{equation}
here $\Vec{a}$ is a column vector whose elements are $a_{i}$, $T$ is correlation matrix with $t_{ij}$ as elements, and $\lambda_{max}$ is a largest eigenvalue of matrix $k= \Vec{a}\Vec{a}^{T}+T T^{T} $. For the case of Bell diagonal states, this expression gets simplified to 
\begin{equation}
    D_{G}(\rho_{BD})= \frac{1}{4}(t_{11}^{2} +t_{22}^{2}+t_{33}^{2} - \max{\{t_{11}^{2},t_{22}^{2},t_{33}^{2}\}}).
\end{equation}
\subsection{Quantum Support Vector Machine}
Quantum support vector machine is a mathematical variant of a classical support vector machine that embodies a hybrid quantum-classical approach. The classical support vector machine is based on a quadratically constrained problem, where the algorithm tries to maximize the distance between the hyperplane, drawn on the edges of different datasets belonging to different classes, while satisfying the constraints \cite{cortes1995support}, given as
\begin{equation}
    L = min(\frac{1}{2}||w||^{2}) - \left(\sum_{i=1}^{d} y_{i}\alpha_{i}K(\Vec{x_{i}},\Vec{x_{j}})+b\right),
\end{equation}
where $x_i$ is the support vector, $w$ is normal to hyperplane, $b$ is offset, $y_i$ is the label of support vector such that $y_i \in \{-1,1\}$, and $K(\Vec{x_{i}},\Vec{x_{j}})$ is a kernel matrix in between two vectors. Classically, the kernel matrix is known as a Gram matrix, whose entries are computed using a dot product between a pair of vectors. Quantum mechanically, the kernel can be simply computed using the inner product or SWAP test \cite{schuld2021supervised} between two different quantum states \cite{havlivcek2019supervised} such that
\begin{equation}
K(\Vec{x_{i}},\Vec{x_{j}})=\abs{\bra{\psi_{\Vec{x_i}}}\ket{\psi_{\Vec{x_j}}}}^2 = tr(\rho_{(x_{i})},\rho_{(x_{j})}),
\label{qsvm}
\end{equation}
where $\Vec{x_i}, \Vec{x_j}$ $\in$ dataset. To compute the SWAP test using a quantum circuit, $\abs{\bra{\psi_{\Vec{x_i}}}\ket{\psi_{\Vec{x_j}}}}^2$ can be further defined as
\begin{equation}
\abs{\bra{\psi_{\Vec{x_i}}}\ket{\psi_{\Vec{x_j}}}}^2 = 
\abs{\bra{0^n}\mathcal{U}^{\dagger}_{\psi(\Vec{x_i})}\mathcal{U}_{\psi({\Vec{x_j}})}\ket{0^n}}^2.
\label{qsvm2}
\end{equation}
From Eq.\eqref{qsvm2}, it becomes evident that kernel values can be feasibly computed using quantum circuits through the evolution of the initial state $\ket{0}$ under the influence of a unitary operator $\mathcal{U}_{\psi(x)}$ and then finally measuring the state in the computational basis. In this study, we use fidelity quantum kernels \cite{thanasilp2022exponential}. Finally, the computed kernel value is passed to the classical support vector machine (CSVM) for classification.

\section{\label{sec:level5}Framework}
With the representation of Bell diagonal states \cite{horodecki2009quantum,PhysRevLett.105.150501}, as shown in Fig. \ref{fig:visualization}, our intent is to train our algorithm on one type of the Werner state and subsequently test it on other types of Werner states, not used during the training. For a comprehensive cross-domain classification and to assess the versatility of the proposed algorithm, in the following subsections, we present an overview of the quantum states utilized for training and testing.

\subsection{\label{sec:level3}Generating dataset}
We now proceed to use quantum machine learning to address the challenges of quantum state classification in two-qubit mixed states using cross-domain classification. In order to facilitate the discussions on the results obtained, in this section, we first describe the generation of generalized two-qubit quantum states with their corresponding labels. The dataset is generated using the generalized two-qubit density matrix, represented in Eq.(\ref{eq:gdm}), where $a_{i}$ and $b_{j}$ are individual polarization vectors and $t_{ij}$ are elements of a correlation matrix- varying the polarization vector and correlation matrix elements, the Werner, Horodecki, and MEMS states are generated.\par
For example, Eq.(\ref{eq:Werner state}) represents a class of Werner states where for $p>\frac{1}{3}$, Werner states are entangled, and for $p \leq \frac{1}{3}$, Werner states are separable. The four types of Werner states are further represented in Fig. \ref{fig:visualization}A for cross-domain classification. \par
Similarly, the Horodecki class of states are entangled for $p>0$, and are expressed as 
\begin{equation}\label{eq:Horodecki state}
    \rho_{H}= p \ket{\psi}\bra{\psi} + (1-p)\ket{00}\bra{00},
\end{equation}
where, $\ket{\psi}$ is one of the Bell states, and $p$ represents probability. Moreover, the MEMS states are a mixture of Bell states with a diagonal separable mixed state \cite{singh2018correlations,wei2003maximal}. The form of MEMS is as follows
\begin{equation}\label{eq:MEMS}
\rho_{MEMS}=\begin{pmatrix}
    q+\frac{\lambda}{2}& 0&0 &\frac{\lambda}{2} \\
    0 & s & 0 & 0\\
    0 & 0 & t & 0\\
    \frac{\lambda}{2}& 0 & 0 & r+ \frac{\lambda}{2}
\end{pmatrix}, 
\end{equation}
where q, r, s, t and $\lambda$ are positive real state parameters and satisfy the condition (q + r + $\lambda$ + s + t)=1. As discussed in Eq.(\ref{eq:bd_state}), we further generate random Bell diagonal states depending on values of $t_{ii}s$ for classifying states as zero and non-zero discord states. Considering the density matrix to be a positive operator, the required necessary constraints on $t_{ii}$s are 
\begin{align*}
1- t_{11} + t_{22} + t_{33}&\geq 0,  &  1+ t_{11} - t_{22} + t_{33}&\geq 0,  \\
1+ t_{11} + t_{22} - t_{33}&\geq 0,  &  1- t_{11} - t_{22} - t_{33}&\geq 0.  
\end{align*}
The states with zero discord belong to one class, and states with non-zero discord belong to another class. Fig. \ref{fig:visualization}B demonstrates the distinction between training and testing domains by restricting the range of $t_{ii}$ from -1 to 0 and from 0 to 1, respectively.  
\begin{figure}
    \centering
    \includegraphics[width=0.5\textwidth]{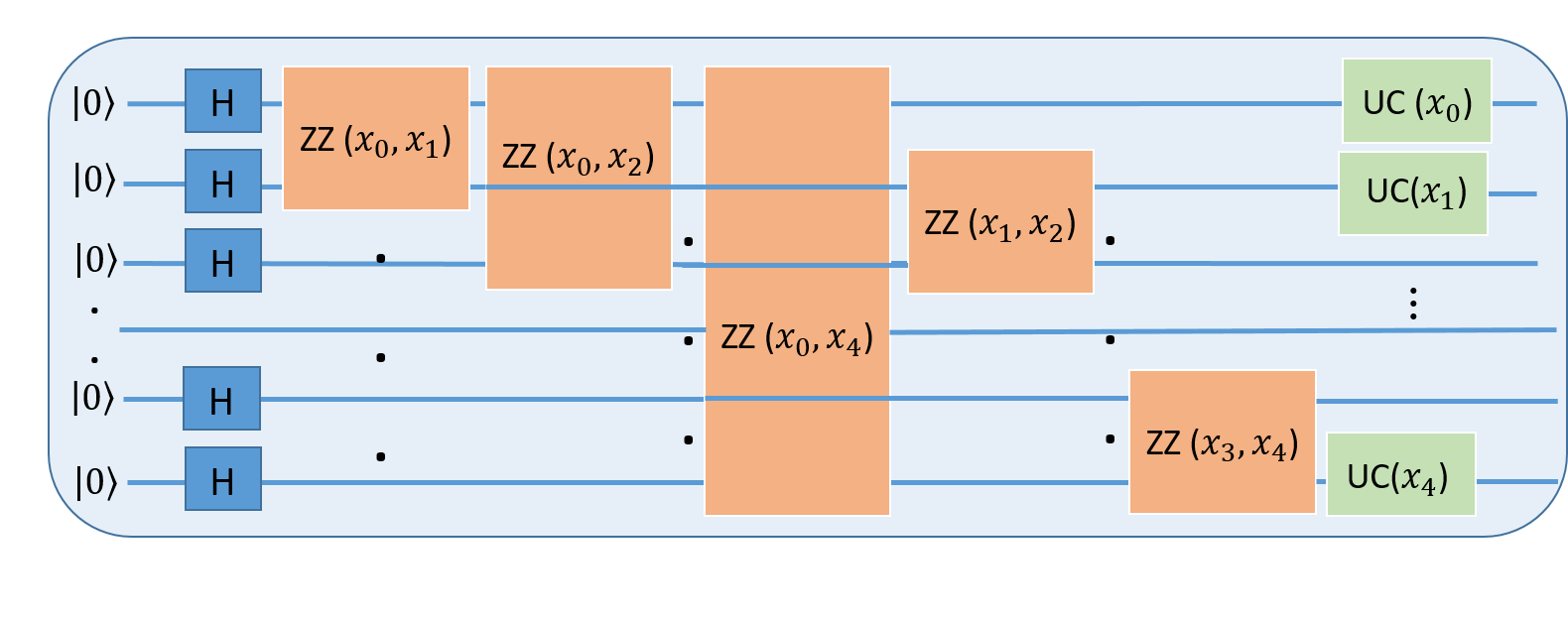}
    \caption{The quantum circuit used in QSVM.}
    \label{fig:circuit}
\end{figure}
\begin{figure}
    \centering
    \includegraphics[width=0.5\textwidth]{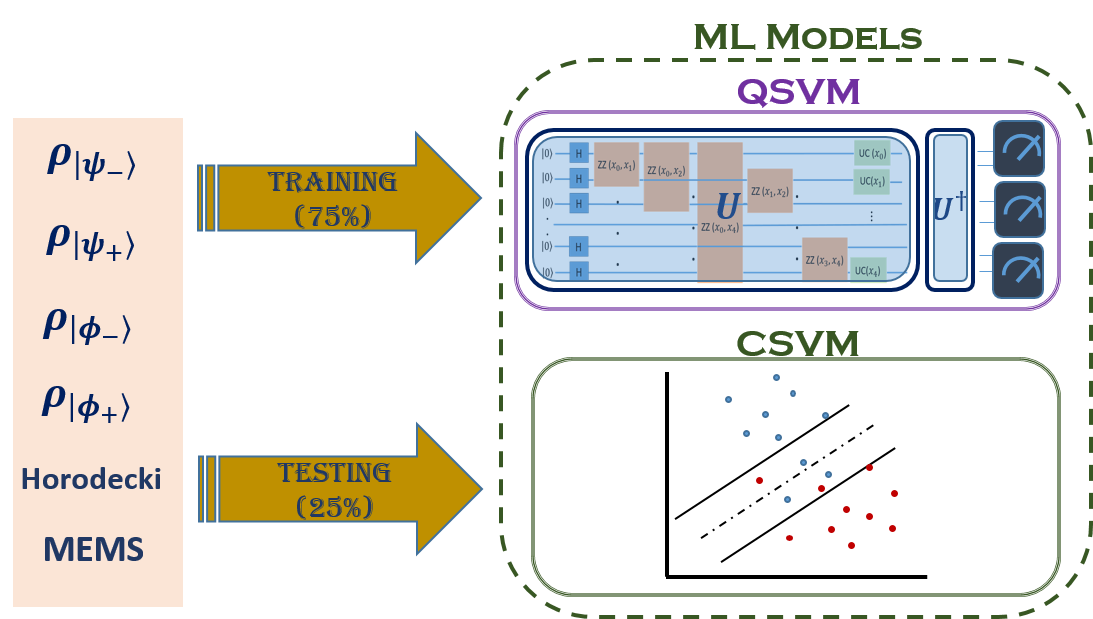}
    \captionof{figure}{A flow chart representing the in-domain classification where we train models (quantum and classical support vector machine) on 75\% of the generated dataset and test models on the rest 25\% of the generated dataset.}
    \label{fig:method1}
\end{figure}

\subsection{\label{sec:level4}Random unitary operations on generated dataset}
For a comprehensive analysis to test the robustness of our model, we extend our approach by generating a collection of random unitary operators to map Werner states to their local unitary transformed forms. The generation of local unitary equivalents of Werner states is accomplished by applying a generalized unitary matrix, given as
\begin{equation}
    U(\theta) = \begin{pmatrix}
        \cos{\theta} & -\sin{\theta}\\
        \sin{\theta} & \cos{\theta}
    \end{pmatrix}.
\end{equation}
To generate the required data for a two-qubit state, we consider a local unitary transformation using the tensor product  ($U(\theta_{1}) \otimes U(\theta_{2})$). The required theta is generated randomly using the Python random module from NumPy \cite{2020NumPy-Array}.
\begin{figure}
    \centering
    \includegraphics[width=0.5\textwidth]{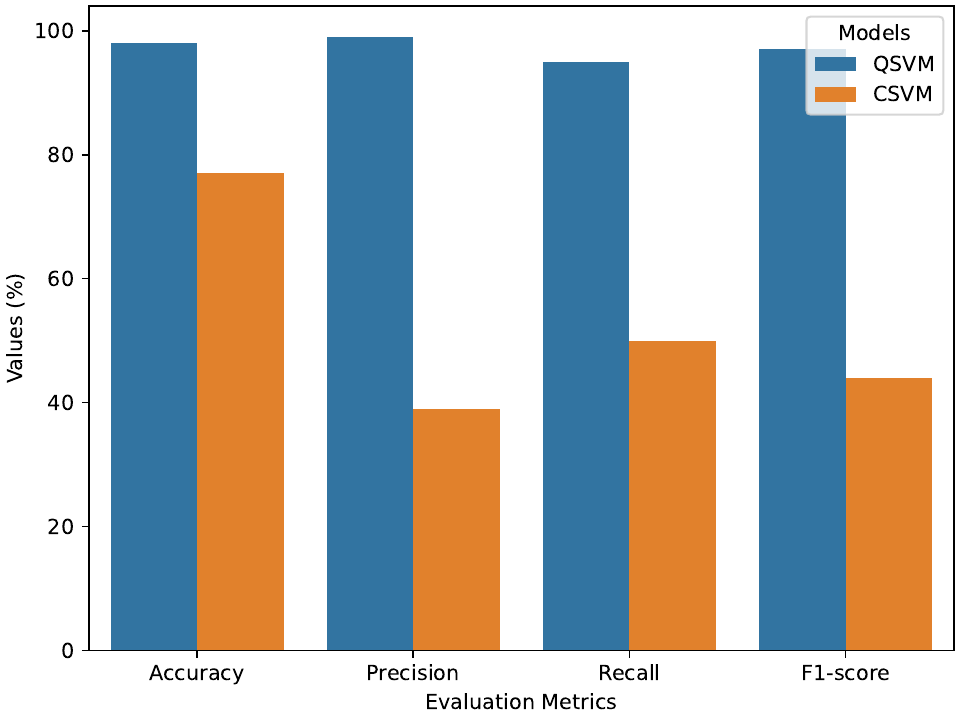}
    \captionof{figure}{A comparison of quantum and classical support vector machines on four evaluation metrics (accuracy, precision, recall, and F1-score) for a combined dataset containing Werner-type, Horodecki, and MEMS states.}
    \label{fig:qsvm_csvm_full}
\end{figure}
\begin{figure}
    \centering
    \includegraphics[width=0.5\textwidth,height=4.5cm]{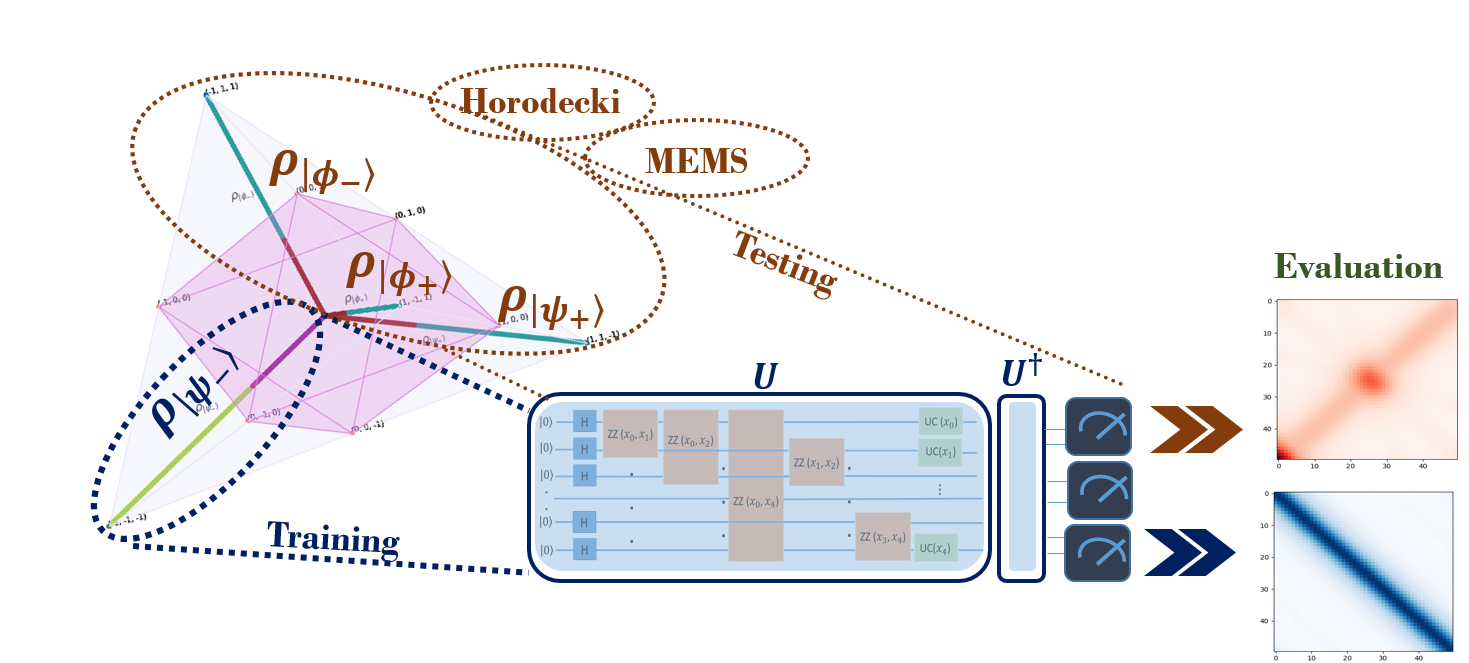}
    \captionof{figure}{A pictorial representation of cross-domain classification where we train our model on $\rho_{\ket{\psi_{-}}}$ and test our model on $\rho_{\ket{\psi_{+}}}$, $\rho_{\ket{\phi_{-}}}$, $\rho_{\ket{\phi_{+}}}$, Horodecki and MEMS states.}
    \label{fig:method2}
\end{figure}
\begin{figure}[hb]
    \centering
    \includegraphics[width=0.5\textwidth,height=4cm]{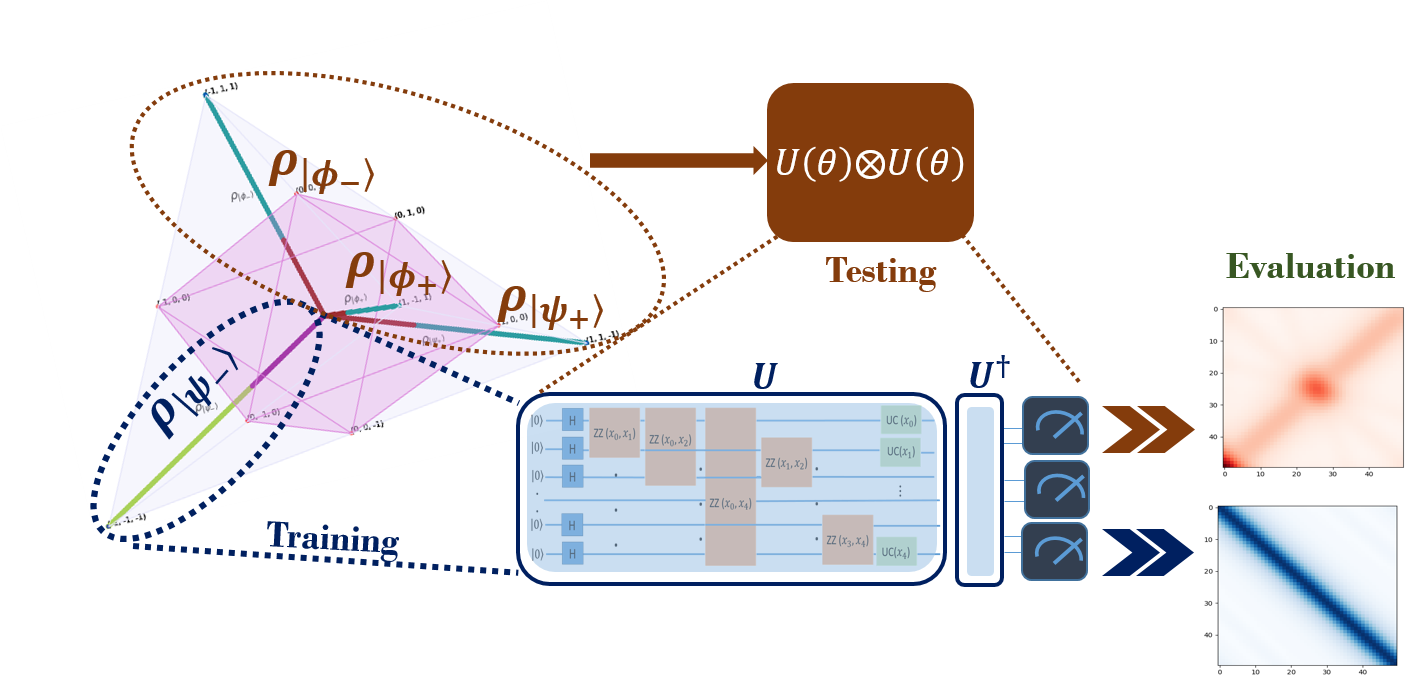}
    \captionof{figure}{The cross-domain classification where we train our model on $\rho_{\ket{\psi_{-}}}$ and test our model on unitary equivalents of  $U \rho_{\ket{\psi_{+}}} U^{\dagger}$,$U \rho_{\ket{\phi_{-}}}U^{\dagger}$ and $U \rho_{\ket{\phi_{+}}} U^{\dagger}$ and similarly for other combinations.}
    \label{fig:method3}
\end{figure}
\section{Analytical Results}
In this section, we analyze effects of utilizing the QSVM (using IBM quantum services \cite{Qiskit}) in comparison to the CSVM (using sklearn library \cite{scikit-learn}) for cross-domain classification of two-qubit mixed states. In the case of QSVM, we use a quantum circuit to compute the distance between quantum states. The mathematical form of the circuit is given as
\begin{equation}
    \begin{aligned}
    \mathcal{U}_{\phi(\Vec{x_i})} &=\\ 
    & exp \left(\sum_{j=0}^{d-1} \sum_{j^{'}=j+1}^{d-1}  Z_{j}Z_{j^{'}} \phi(x_j) \phi(x_{j^{'}}) + \sum_{j=0}^{d-1} UC_{j}\phi(x_{j})\right).
    \end{aligned}
\end{equation}
\begin{figure*}[htpb]
\centering
\captionsetup[subfigure]{labelformat=empty}
\subfloat[A]{\includegraphics[width=0.6\textwidth]{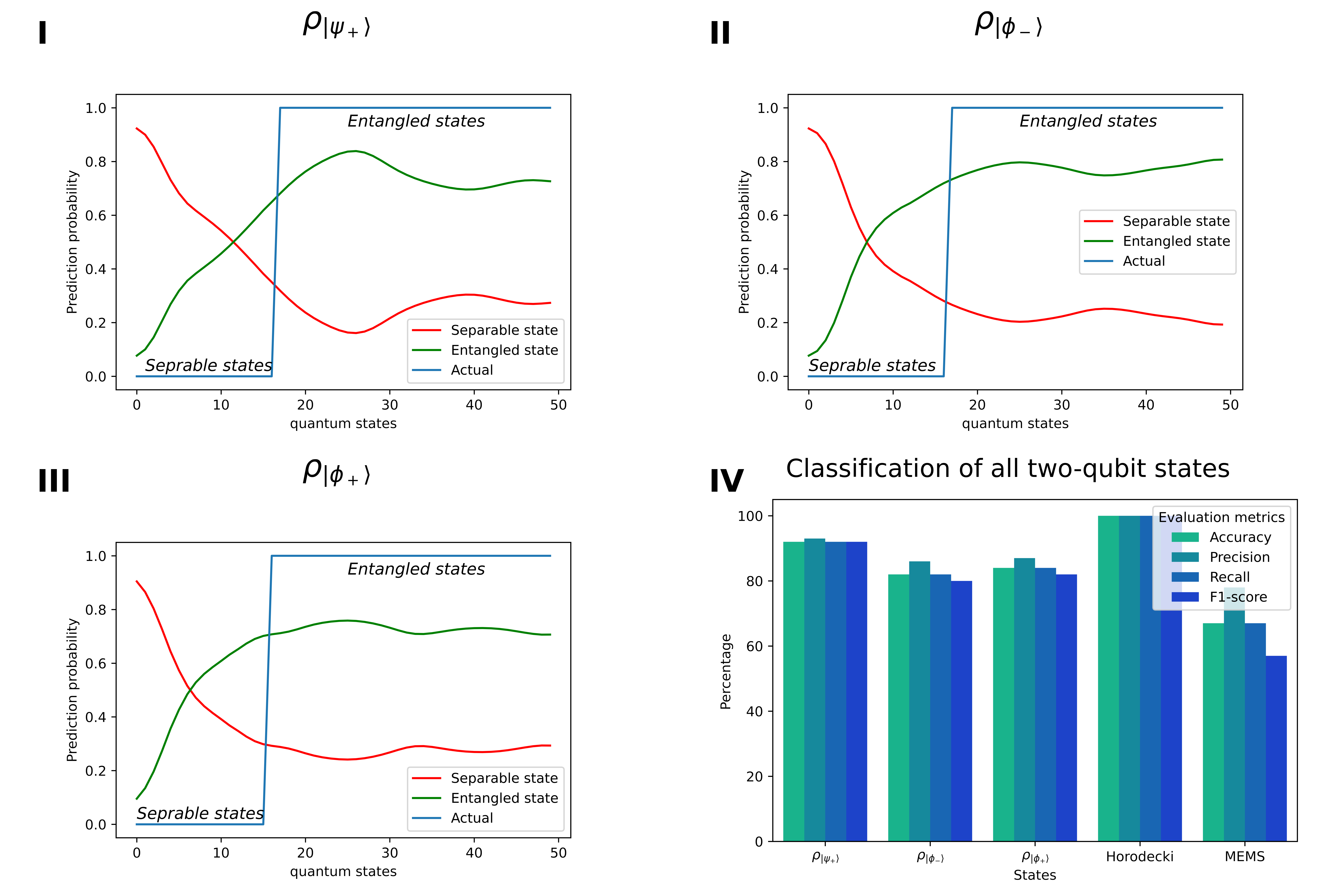}}
\subfloat[B]{\includegraphics[width=0.45\textwidth]{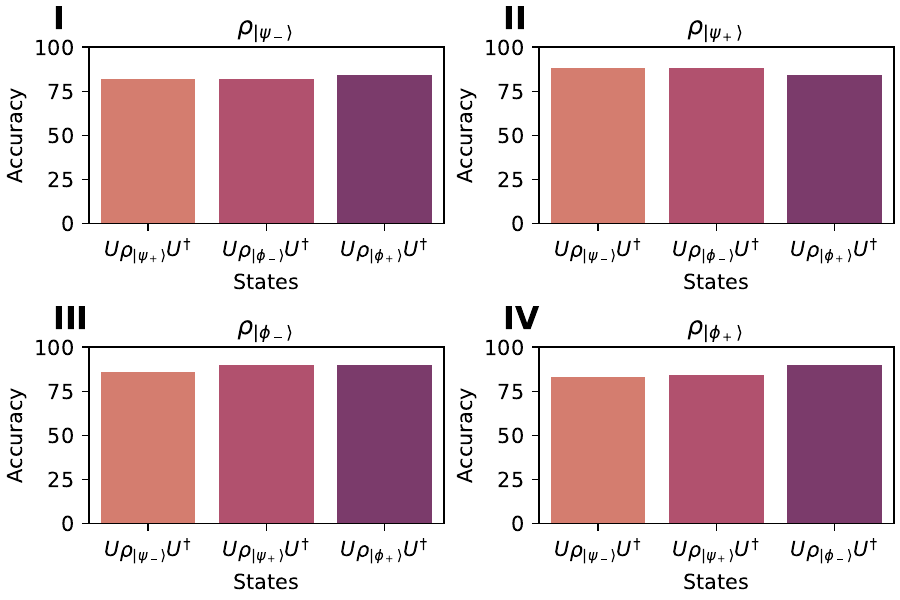}}
\captionof{figure}{A-I, A-II, and A-III represent the model being trained on $\rho_{\ket{\psi_{-}}}$ and tested on $\rho_{\ket{\psi_{+}}}$, $\rho_{\ket{\phi_{-}}}$, and $\rho_{\ket{\phi_{+}}}$, respectively- the results demonstrate prediction probabilities corresponding to each testing quantum states. A-IV represents the efficiency of our model for cross-domain classification using four different evaluation metrics. B) Accuracy of our model for classifying unitary equivalent of Werner type of states.}
\label{fig:werner_states}
\end{figure*}
As per the above equation, the circuit consists of ZZ and UC gates, where the UC gate is a combination of $R_{x}$ and $R_y$ gates, and $\phi(x_j)$ and $ \phi(x_{j^{'}})$ are the elements of the generated density matrices. The circuit is chosen after performing an optimization based on repetitions of the gates and a number of measurements, and in comparison to other circuits such as instantaneous quantum polynomial (IQP) and Hamiltonian circuits \cite{havlivcek2019supervised, huang2021power}. Fig. \ref{fig:circuit} shows the circuit used to implement QSVM. For CSVM, we optimize the model based on linear and non-linear kernels in addition to gamma and C, which are hyperparameters \cite{vert2004primer}. \par
In accordance with the conventional methods of in-domain classification, we first train the quantum and classical models with $75$$\%$  of the whole dataset and test the models on the rest $25\% $ of the data as represented in Fig. \ref{fig:method1}. Further, Fig. \ref{fig:qsvm_csvm_full} demonstrates the classification efficiency of the respective models in terms of accuracy, precision, recall, and F1-score. Our results indicate an accuracy of $97.72 \%$ using QSVM compared to the accuracy of $77.27 \%$ using CSVM. Moreover, the precision value for CSVM is only $39\%$, indicating that a significant number of states are predicted as false positive, i.e., false entangled states instead of separable states. However, the recall value for the classical model is relatively better than the precision value, specifying that the predicted false negative or false separable states are lesser in number. These observations clearly indicate that the CSVM exhibits type-I error \cite{banerjee2009hypothesis}. On the other hand, our QSVM model shows significant results in accuracy as well as other evaluation metrics (precision, recall, and F1-score). The results indicate that QSVM is more reliable when classifying the quantum states as entangled or separable. \par
To comprehensively analyze the models for cross-domain classification, we proceed strategically to address the entanglement versus separability paradigm in the following subsections.

\newcommand{\ra}[1]{\renewcommand{\arraystretch}{#1}}

\begin{table*}[!ht]
\centering
\ra{1.5}
\begin{tabular}{|c|ccc|ccc|ccc|ccc|ccc|}
\hline
Training states & \multicolumn{3}{c|}{$\rho_{\ket{\psi_{-}}}$} & \multicolumn{3}{c|}{$\rho_{\ket{\psi_{+}}}$} & \multicolumn{3}{c|}{$\rho_{\ket{\phi_{-}}}$} & \multicolumn{3}{c|}{$\rho_{\ket{\phi_{+}}}$} \\
\hline
Testing states & $\rho_{\ket{\psi_{+}}}$ & $\rho_{\ket{\phi_{+}}}$ & $\rho_{\ket{\phi_{-}}}$ & $\rho_{\ket{\psi_{-}}}$ & $\rho_{\ket{\phi_{+}}}$ & $\rho_{\ket{\phi_{-}}}$ & $\rho_{\ket{\psi_{+}}}$ & $\rho_{\ket{\psi_{-}}}$ & $\rho_{\ket{\phi_{+}}}$  & $\rho_{\ket{\psi_{+}}}$ & $\rho_{\ket{\psi_{-}}}$ & $\rho_{\ket{\phi_{-}}}$  \\
\hline
Accuracy &  92\% & 84\% & 82\%  & 94\% & 84\% & 86\%  & 86\% & 82\% & 94\% &82\% & 84\% & 92\% \\
\hline
\end{tabular}
\caption{A detailed evaluation of Werner type states for cross-domain classification.}
\label{tab:all_werner_states}
\end{table*}

\subsection{\label{subsec:level5.1} Cross-domain classification using QSVM}
In this subsection, we proceed to analyze the efficacy of the proposed QSVM for cross-domain classification for characterizing entangled and separable states, which are spatially different from the training states in the Hilbert space. For this, we utilize the Werner states depicted in Fig. \ref{fig:visualization}A. We inquisitively train the QSVM on a single type of Werner state ($\rho_{\ket{\psi_{-}}}$) and subsequently test the trained model with different Werner states ($\rho_{\ket{\psi_{+}}}$, $\rho_{\ket{\phi_{-}}}$ and $\rho_{\ket{\phi_{+}}}$), as shown in Fig. \ref{fig:method2}. Clearly, Fig. \ref{fig:werner_states}A demonstrates the advantages of using QSVM for classifying the Werner, Horodecki, and MEMS states as entangled or separable. For example, Figs. \ref{fig:werner_states}A- I, II and III represent the QSVM predictions corresponding to $\rho_{\ket{\psi_{+}}}$, $\rho_{\ket{\phi_{-}}}$ and $\rho_{\ket{\phi_{+}}}$, respectively when the model is trained using a Hilbert space occupied by the $\rho_{\ket{\psi_{-}}}$ states (refer to Fig. \ref{fig:visualization}A). Fig. \ref{fig:werner_states}A-I confirms that our model has a high prediction probability to accurately classify a separable (entangled) state if the actual testing state is separable (entangled). Of course, there are a few separable states, which are misclassified as entangled states as the value of the state parameter $p$ approaches $\frac{1}{3}$. Similar observations also result from figures \ref{fig:werner_states}A-II and  \ref{fig:werner_states}A-III. Therefore, our results show that the proposed model can predict the separable and entangled states with a significant probability. The same can be inferred from plots of evaluation metrics, i.e., accuracy, precision, recall, and F1-score for Werner type of states as represented in Fig. \ref{fig:werner_states}A-IV. We further train our model on other types of Werner states for cross-domain classification and achieve similar results. In fact, training the model on $\rho_{\ket{\psi_{+}}}$ or $\rho_{\ket{\phi_{-}}}$ results in an increased accuracy by $2\%$ for classification of $\rho_{\ket{\psi_{-}}}$ or $\rho_{\ket{\phi_{+}}}$. Table \ref{tab:all_werner_states} summarizes the results for the cross-domain classification of Werner states. \par
For a comprehensive analysis of two-qubit states using the cross-domain classification strategy, we further evaluate our model on Horodecki and MEMS states while training the model on the Werner state ($\rho_{\ket{\psi_{-}}}$). Our results show that the proposed model is highly competent even for the cross-domain classification of Horodecki and MEMS states. Surprisingly, for Horodecki states, QSVM shows a $100\%$ success on all evaluation metrics. For MEMS states, the QSVM shows a moderate accuracy compared to Werner and Horodecki states. This can be attributed to the significant difference between the $\rho_{\ket{\psi_{-}}}$ state and MEMS states. Nevertheless, the accuracy is still significant for the cross-domain classification. Hence, QSVM successfully performs the cross-domain classification and effectively addresses the entanglement versus separability problem. 
\subsection{\label{subsec:level5.2} Analysing robustness of QSVM}
In order to further ascertain the predictive capacity of the QSVM on Werner states, we train the model on one type of Werner state and then test the model on the local unitary equivalents of other types of Werner states. Fig. \ref{fig:method3} is a schematic representation of our approach. Fig. \ref{fig:werner_states}B represents the accuracy corresponding to the cross-domain classification of different types of Werner states. Our analysis reveals that the model consistently achieves adequate accuracy in predicting entangled and separable states. 

\begin{figure*}[htpb]
\centering
\captionsetup[subfigure]{labelformat=empty}
\subfloat[A. Results of cross-domain classification on Bell diagonal states]{\includegraphics[width=0.45\textwidth]{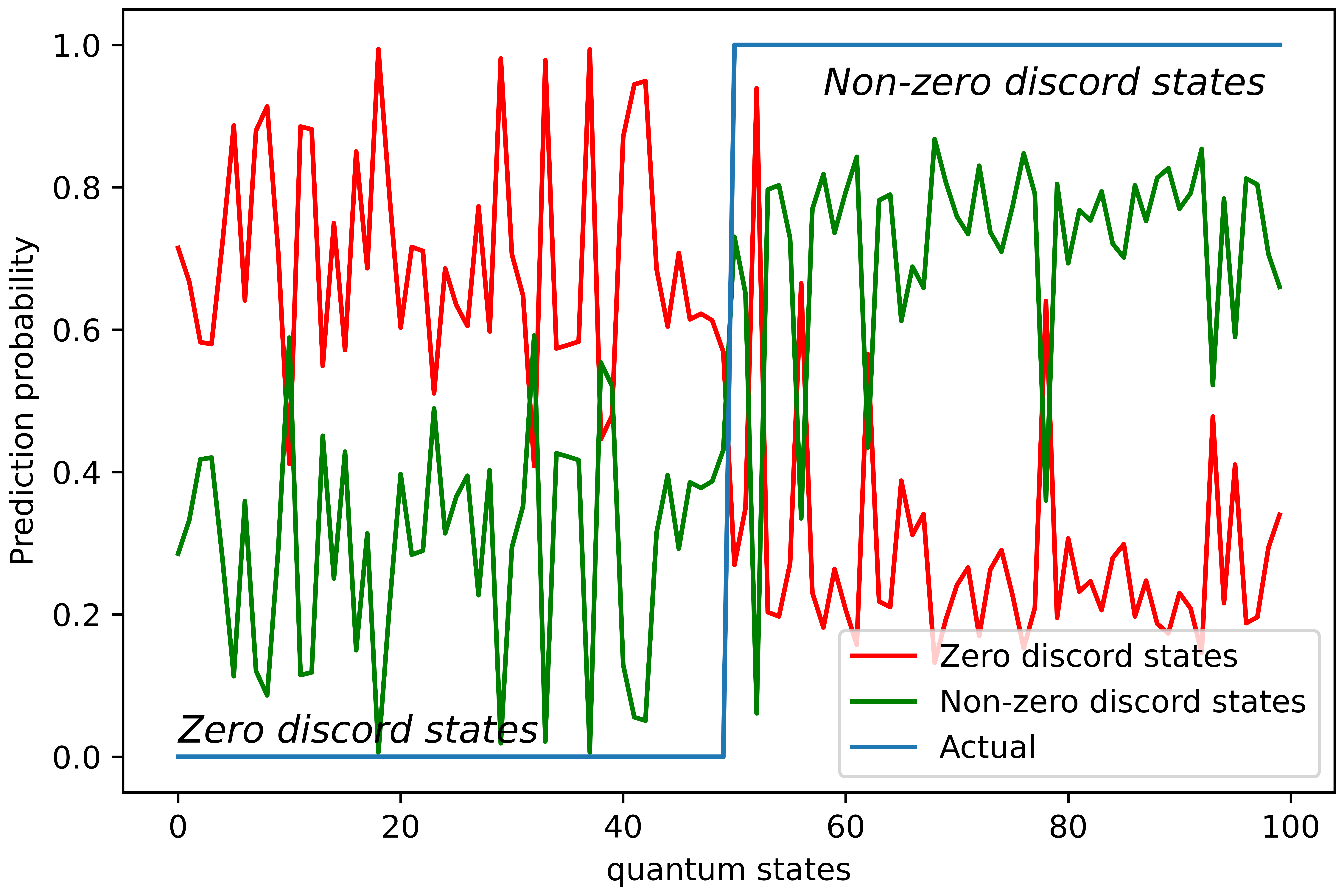}}
\subfloat[B. Predictions corresponding to Werner states, trained on Bell diagonal states ($-1 <t_{ii} < 0 $)]{\includegraphics[width=0.5\textwidth]{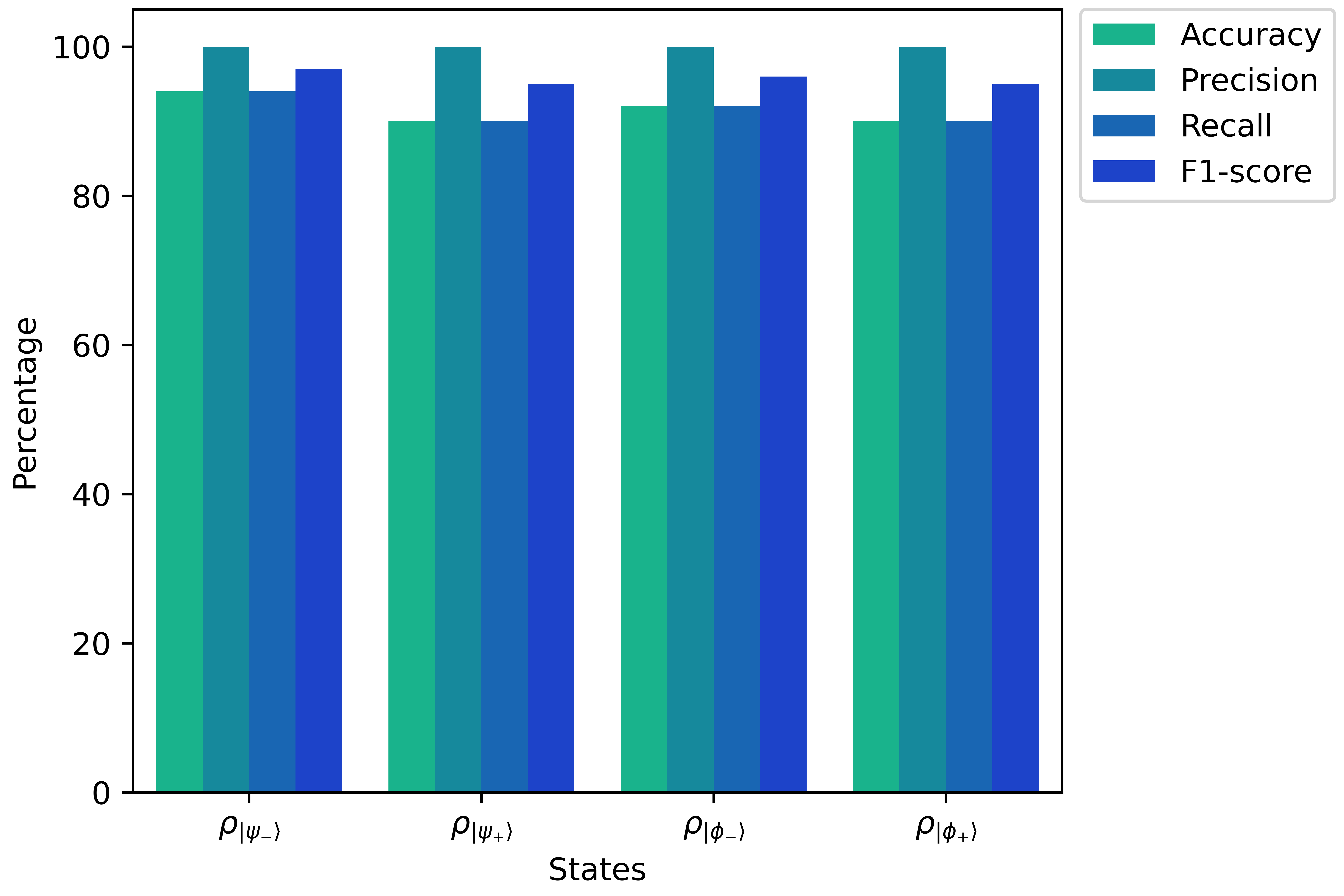}}
\captionof{figure}{For cross-domain classification of zero and non-zero discord states (Fig. \ref{fig:visualization}B), A) demonstrates classification results where the blue line represents the actual class of testing states. Similarly, red and green lines represent the prediction probabilities corresponding to the classification of states in zero and non-zero discord states. B) shows the classification results of Werner states based on different evaluation metrics.}
\label{fig:discord_states}
\end{figure*}

\begin{table*}[!t]
    \centering
    \begin{tabular}{|c|c|c|c|}
    \hline
        Sample Size & $N_{h}=0$ & $N_{h}=50$ & $N_{h}=100$  \\
        \hline
         $5*10^{3}$ & Predicted as all separable states &Predicted as all separable states & Predicted as all separable states \\
         \hline
         $5*10^{4}$ & Predicted as all separable states & Predicted as all separable states & Predicted as all separable states\\
         \hline
    \end{tabular}
    \caption{Analysis of NN with increased number of neurons in the hidden layer ($N_h$) and sample size.}
    \label{tab:ann}
\end{table*}

\section{Analysing non-classical correlation using QSVM}
In this section, we evaluate our model for predicting non-classical correlations existing in two-qubit mixed states. As discussed in section \ref{sec:level3}, the training and testing states are generated by restricting the range of $t_{ii}$ (Eq.\eqref{eq:bd_state}). Fig. \ref{fig:discord_states}A demonstrates the prediction probabilities for classifying Bell diagonal states into zero or non-zero discord states. Further, the accuracy and precision of classifying the Bell diagonal states are $73\%$ and $82\%$, respectively. In addition to the classification of Bell-diagonal states, we further analyze Werner-type states for non-classical correlations while still training our model on Bell-diagonal states. Fig. \ref{fig:discord_states}B exhibits the excellent potential of our model for all evaluation metrics. 

\section{Analysis of classical machine learning models}
For cross-domain classification (Fig. \ref{fig:method2}) of quantum states, we also evaluate the prediction capability of classical machine learning algorithms- CSVM and neural networks (NN). In the case of CSVM, we observe that the classification of Werner states as entangled or separable resulted in poor accuracy. The model predicts each testing state as separable using the linear or non-linear kernel with optimized C and gamma parameters. Similarly, NN also predicts all testing states as separable, with a loss of 0.044 on the training dataset. The predictions for the testing dataset using NN remain the same even with increasing epochs and the number of neurons in the hidden layer. Table \ref{tab:ann} represents a summary of NN results corresponding to the number of neurons in hidden layers and the sample size. We observe similar results using CSVM and NN for the classification of zero and non-zero discord states.

\section{Conclusion}
This study delved into the efficacy of employing a quantum support vector machine (QSVM) to address the entanglement versus separability paradigm in two-qubit mixed states using cross-domain classification. Using the proposed algorithm, we efficiently classified two-qubit mixed states belonging to different Hilbert spaces. The results obtained here testify to our algorithm's efficacy and robustness in achieving quantum advantages compared to conventional methods (CSVM and NN) for cross-domain classification of quantum states. Interestingly, for the classification of Horodecki states, our results have shown $100\%$ success on all evaluation metrics. For Werner-type states, we have trained our model on $\rho_{\ket{\psi_{-}}}$ state and tested on other types of Werner states. The reason to use $\rho_{\ket{\psi_{-}}}$ states as training set aligns with the antisymmetric properties of the $\ket{\psi_{-}}$ which are different from other Bell states. Moreover, we also trained our model on different types of Werner states and the results for cross-domain classification remain intact. In fact, in certain cases, we observed an increase in accuracy by $2\%$ for classifying states as separable or entangled. For example, we achieved an accuracy of $94\%$ when we trained our model on $\rho_{\ket{\psi_{+}}}$ and tested our model on $\rho_{\ket{\psi_{-}}}$. In addition to the separability and entanglement problem, we further addressed the identification of mixed states exhibiting non-classical correlations. \par
Therefore, we conclusively demonstrated the potential of the QSVM to detect entanglement and non-classical correlations in cross-domain regimes, which is not the case with CSVM and NN. Clearly, for two-qubit mixed-state classification, the effectiveness of classical machine learning algorithms remains bounded within the realms of in-domain classification. It will be interesting to extend our study to a multi-qubit scenario.

\section{Data Availbility}
The data set involving two-qubit mixed states of different classes is generated using Quantum Toolbox in Python (QuTiP) \cite{JOHANSSON20121760,JOHANSSON20131234}. The data is available in the  \href{https://github.com/viveksabale1998/Quantum-States-Data-Set.git}{GitHub} repository.


\bibliography{apssamp}

\end{document}